\documentclass[11pt,a4paper]{article}
\usepackage{jcappub}
\usepackage{amsmath,amssymb,graphics,epsfig,subfigure}
\usepackage{color}

\title{Observing the shadow of Einstein-Maxwell-Dilaton-Axion black hole}

\author{Shao-Wen Wei,}
\author{Yu-Xiao Liu}
\affiliation{Institution,\\Theoretical Physics, Lanzhou University, Lanzhou 730000, People's Republic of China}

\emailAdd{weishw@lzu.edu.cn, liuyx@lzu.edu.cn}

\abstract{
In this paper, the shadows cast by Einstein-Maxwell-Dilaton-Axion black hole and naked singularity are studied. The shadow of a rotating black hole is found to be a dark zone covered by a deformed circle. For a fixed value of the spin $a$, the size of the shadow decreases with the dilaton parameter $b$. The distortion of the shadow monotonically increases with $b$ and takes its maximal when the black hole approaches to the extremal case. Due to the optical properties, the area of the black hole shadow is supposed to equal to the high-energy absorption cross section. Based on this assumption, the energy emission rate is investigated. For a naked singularity, the shadow has a dark arc and a dark spot or straight, and the corresponding observables are obtained. These results show that there is a significant effect of the spin $a$ and dilaton parameter $b$ on these shadows. Moreover, we examine the observables of the shadow cast by the supermassive black hole at the center of the Milky Way, which is very useful for us to probe the nature of the black hole through the astronomical observations in the near future.}


\begin{document}


\maketitle

\section{Introduction}

It is widely believed that there exist supermassive black holes in the
centers of many galaxies. Due to the rotation of galaxies, these black hole are generally thought to possess a spin parameter. Then, there is a great interest to probe the nature of the black holes, i.e., the mass and spin of them.

In general, the mass of a black hole can be estimated with the Newtonian
orbital motion of the stars surrounding the black hole \cite{Eisenhauer}. And the black hole spin can be measured with the methods based on the thermal continuum emission of accretion disks and relativistically broadened iron lines \cite{Tanaka}. These two techniques using X-ray measurements have been widely applied to different sources. However, for sources such as Sgr A$^*$, the two techniques are not applicable. Then an alternate method may be the observation of black hole shadows. Black hole shadow is the optical appearance casted by a black hole in the sky. And near the shadow, the strong gravitational lensing may be very obvious. For an example, there will be a large Einstein ring and two infinite series of concentric relativistic Einstein rings near a nonrotating black hole, and only one or two bright images for a rotating black hole. These are corresponded to the strong gravitational lensing, which is well studied when the photon passes near the photon sphere of a black hole \cite{Darwin,BozzaCapozziello,Eiroa,Virbhadra,Chen,SarkarBhadra,Whisker,Bhadra,
Vazquez,Gyulchev75,Nandi,wei}. Compared with relativistic images, the shadow of a black hole is a two-dimensional dark zone seen from the observer, so it can be easily observed. For a nonrotating black hole, its shadow is a perfect circle. While for the rotating case, it has an elongated shape in the direction of the rotation axis due to the dragging effect \cite{Bardeen,Chandrasekhar}. And this subject has been studied in the last few years \cite{Falcke,Vries,Takahashi,Hioki,Bambi,Kraniotis} because that the observations may be obtained in the near future \cite{Zakharov}. Therefore, the investigation of the shadow is very useful for measuring the nature of the black hole.

Optical properties and apparent shape of the rotating black holes have been investigated in Refs. \cite{Bozzagrg,Schee}. In Ref. \cite{Schee}, the authors studied the optical phenomena in a braneworld Kerr black hole, and then applied to Sgr A$^{*}$ supermassive black hole located at the center of the Milky Way. Through constructing the observables, the shadow of a Kerr black hole or a Kerr naked singularity was examined in Ref. \cite{Maeda}. The authors analyzed the effect of the spin parameter and inclination angle on the apparent shape of the shadow, and suggested that the spin parameter and inclination angle of a Kerr black hole or a naked singularity can be determined from the observation. This was also extended to the rotating black hole in extended Chern-Simons modified gravity \cite{Amarilla}, the rotating braneworld black hole \cite{AmarillaEiroa}, the multi-Black Holes \cite{YumotoNitta}, and as well as to the Kaluza-Klein rotating dilaton black hole \cite{Amarilla13} and the rotating traversable wormholes or black holes \cite{Nedkova}. These results show that, beside the spin, other parameters (i.e, coupling constant and tidal charge) also affect the shadow of a black hole. So, the observation of shadow also provides a possible way to probe the parameters of a black hole.

In this paper, with the observables proposed in \cite{Maeda}, we will study the apparent shape of the shadow for the Einstein-Maxwell-Dilaton-Axion (EMDA) black hole and naked singularity \cite{Garcia}, respectively. For an example, we assume that, near the supermassive black hole Sgr A$^{*}$, the spacetime can be described by the EMDA black hole metric, and far away from it, the spacetime is flat as we expect. This means that the influence of the black hole is local. However, the photon emitted from the vicinity of the black hole carries the information of it, which can be tested at infinity. Under this assumption, the observables and angular radius are obtained. The results show that, the resolution of 1 $\mu$as will be enough to extract the information of the dilaton parameter $b$ from further observations, while for the spin $a$, the resolution of less than 1 $\mu$as is needed. These are wished to be observed using the very-long baseline interferometry (VLBI) in the near future \cite{w0,w1}.

The paper is structured as follows. In Sec. \ref{geodesic}, we give a brief review of the null geodesics for the EMDA black hole. In Sec. \ref{Circular}, the circular photon orbits are studied. In Sec. \ref{BHS}, we investigate apparent shape of the shadow cast by the EMDA black hole and the corresponding observalbes are obtained. The energy emission rate is studied through the observalbes. The shadow for a naked singularity is examined in Sec. \ref{NSS}. In the final section, we discuss these results and apply to the supermassive black hole at the center of the Milky Way.

\section{Null geodesics}
\label{geodesic}

\begin{figure*}
\begin{center}
\includegraphics[width=6.5cm]{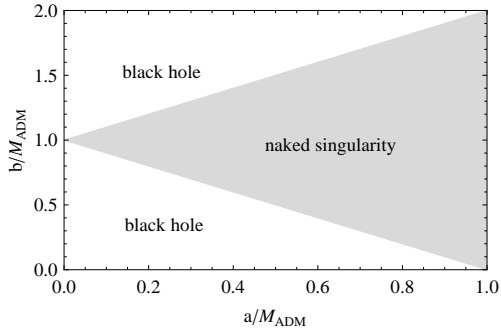}
\end{center}
\caption{The range space of the parameters $a$ and $b$ describing a black hole and a naked singularity.} \label{BHR}
\end{figure*}

The EMDA black hole is described by the following action,
\begin{eqnarray}
 S=\int d^{4}x\sqrt{-g}(R-2\partial_{\mu}\varphi\partial^{\mu}\varphi
   -\frac{1}{2}e^{4\phi}\partial_{\mu}\kappa\partial^{\mu}\kappa
   -e^{-2\phi}F_{\mu\nu}F^{\mu\nu}
   -\kappa F_{\mu\nu}\check{F}^{\mu\nu}),
\end{eqnarray}
where $R$ is the scalar Riemann curvature, $F_{\mu\nu}$ is the electromagnetic antisymmetric tensor field with $\check{F}_{\mu\nu}$ its dual. $\varphi$ and $\kappa$ are the dilaton scalar field and axion field, respectively. From this action, the EMDA black hole can be obtained \cite{Garcia}:
\begin{eqnarray}
 ds^{2}=-\frac{\Delta-a^{2}\sin^{2}\theta}{\Sigma}dt^{2}
        &-&\frac{2a\sin^{2}\theta(\Xi-\Delta)}{\Sigma}dtd\phi
        +\frac{\Sigma}{\Delta}dr^{2}
        +\Sigma d\theta^{2}\nonumber\\
        &+&\frac{\Xi^{2}-\Delta a^{2}\sin^{2}\theta}{\Sigma}\sin^{2}
          \theta d\phi^{2},\label{metric}
\end{eqnarray}
with the metric functions given by
\begin{eqnarray}
 \Sigma&=&r^{2}+2br+a^{2}\cos^{2}\theta,\\
 \Delta&=&r^{2}-2Mr+a^{2},\\
 \Xi&=&r^{2}+2br+a^{2}.
\end{eqnarray}
The parameters $a$ and $b$ represent the spin and dilaton parameters per unit mass of the black hole, respectively. The ADM mass of the black hole is $M_{ADM}=M+b$. When $b=0$, this metric will reduce to the Kerr black hole. The horizons are determined by the equation $\Delta=0$, in terms of the ADM mass,
\begin{eqnarray}
 r_{\pm}=(M_{ADM}-b)\pm\sqrt{(M_{ADM}-b)^{2}-a^{2}}.\label{horizon}
\end{eqnarray}
Here, $r_{+}$ denotes the outer horizon and $r_{-}$ the inner horizon. It is clear that this solution describes a nonextremal black hole for $r_{+}> r_{-}$. When $r_{+}=r_{-}$, one will obtain an extremal black hole. And a naked singularity will appear when $r_{-}>r_{+}$. We show in Fig. \ref{BHR} the range space of the parameters $a$ and $b$ describing a black hole and a naked singularity. In what follows, for simplicity, we adimensionalize all quantities with the ADM mass of the black hole, which is equivalent to put $M_{ADM}=1$ (or $M=1-b$) in all equations.

In general, there are two cases when a photon passes a black hole from infinity. The first case is for the photon with large orbital angular momentum. It will turn back at some turning points and be observed by the observer located at infinity. While the photon with small orbital angular momentum will fall into the black hole without reaching the observer. Thus, there exists a dark zone in the sky called the shadow, which boundary is determined by the two cases. Through studying the geodesic of a photon in the black hole background, one is allowed to obtain the apparent shape of the shadow.

The geodesic for this black hole is determined by the following Hamilton-Jacobi equation
\begin{eqnarray}
 \frac{\partial S}{\partial\lambda}
    =-\frac{1}{2}g^{\mu\nu}\frac{\partial S}{\partial x^{\mu}}
      \frac{\partial S}{\partial x^{\nu}},\label{jacobiequation}
\end{eqnarray}
where $\lambda$ is an affine parameter along the geodesics, and $S$ is the Jacobi action. For this black hole background (\ref{metric}), the Jacobi action $S$ can be separated as
\begin{eqnarray}
 S=\frac{1}{2}m^{2}\lambda-Et+L+S_{r}(r)+S_{\theta}(\theta),\label{action}
\end{eqnarray}
where $S_{r}$ and $S_{\theta}$ are functions of $r$ and $\theta$, respectively. The constants $m$, $E$, and $L$ are, respectively, the particle's mass, energy and angular momentum with respect to the rotation axis. For the photon, we have $m^{2}=0$. Plunging (\ref{action}) into (\ref{jacobiequation}), we can obtain the null geodesics for a photon,
\begin{eqnarray}
 \Sigma\frac{dt}{d\lambda}&=&\frac{(a^{2}+2br+r^{2})(a^{2}E-aL+(r+2b)Er)}{\Delta}
   +aL-a^{2}E\sin^{2}\theta,\label{tr}\\
 \Sigma\frac{d\phi}{d\lambda}&=&\frac{a(a^{2}E-aL+E(r^{2}+2br))}{\Delta}
    +L\csc^{2}\theta-aE,
\end{eqnarray}
\begin{eqnarray}
 \Sigma\frac{dr}{d\lambda}&=&\pm \sqrt{R},\label{radial}\\
 \Sigma\frac{d\theta}{d\lambda}&=&\pm\sqrt{\Theta},\label{thetar}
\end{eqnarray}
where
\begin{eqnarray}
 R&=&-\Delta(K+(L-aE)^{2})+(a^{2}E-aL+(r+2b)Er)^{2},\\
 \Theta&=&K+\cos^{2}\theta(a^{2}E^{2}-\frac{L^{2}}{\sin^{2}\theta}),
\end{eqnarray}
with $K$ is a constant of separation. These equations determine
the propagation of light in the spacetime of the EMDA black hole. When $b=0$, it is just the null geodesic for the Kerr black hole.

\section{Circular photon orbits}
\label{Circular}

In order to obtain the boundary of the shadow of the black hole, we need to study the radial motion (\ref{radial}). First, we rewrite it as
\begin{eqnarray}
 \Sigma^{2}\dot{r}^{2}+V_{eff}=0.
\end{eqnarray}
The effective potential $V_{eff}$ reads
\begin{eqnarray}
 V_{eff}/E^{2}=(a^{2}+(r+2b)r-a\xi)^{2}-(r^{2}+a^{2}+2r(b-1))(\eta+(\xi-a)^{2}),
\end{eqnarray}
where $\xi=L/E$, $\eta=K/E^{2}$. The boundary of the shadow is mainly determined by the circular photon orbit, which satisfies the following conditions
\begin{eqnarray}
 V_{eff}=0, \quad \partial_{r} V_{eff}=0.\label{condition}
\end{eqnarray}
First, we consider the nonrotating case ($a=0$), for which the solution of (\ref{condition}) is
\begin{eqnarray}
 r_{0}=\frac{1}{2}(3-3b+\sqrt{b^{2}-10b+9}).
\end{eqnarray}
As expected, when $b=0$, this just the photon sphere of the Schwarzschild black hole, $r_{0}=3$. At the same time, the parameters $\xi$ and $\eta$ satisfy
\begin{eqnarray}
 \xi^{2}+\eta=\frac{r_{0}^{3}+4br_{0}^{2}+4b^{2}r_{0}}{r_{0}+2b-2}.\label{nonxieta}
\end{eqnarray}
For the rotating case $(a\neq 0)$, we have
\begin{eqnarray}
 \xi&=&-\frac{(r+b+1)a^{2}+r(\Delta-a^{2}+(b-1)(r+2b))}{a(r+b-1)},\\
 \eta&=&r^{2}\frac{4a^{2}(r+b)-(\Delta-a^{2}+(b-1)(r+2b))^{2}}{a^{2}(r+b-1)^{2}}.
\end{eqnarray}
When $b=0$, the result for the Kerr black hole will be obtained \cite{Maeda}. Through analyzing these equations, the shadow of the EMDA black hole can be determined.

\section{Black hole shadow}
\label{BHS}

In this section, we would like to study the black hole shadow, where the parameter $b$ is limited in the ranges $b\leq 1-a$ or $b\geq 1+a$.

In order to study the shadow of the EMDA black hole, we introduce the celestial coordinates
\begin{eqnarray}
 \alpha&=&\lim_{r\rightarrow \infty}
   \bigg(-r^{2}\sin\theta_{0}\frac{d\phi}{dr}
      \bigg|_{\theta\rightarrow \theta_{0}}\bigg)
     =-\xi\csc\theta_{0},\label{alpha}\\
 \beta&=&\lim_{r\rightarrow \infty}
   \bigg(r^{2}\frac{d\theta}{dr}\bigg|_{\theta\rightarrow \theta_{0}}\bigg)
     =\pm\sqrt{\eta+a^{2}\cos^{2}\theta_{0}-\xi^{2}\cot^{2}\theta_{0}},\label{beta}
\end{eqnarray}
where we have used Eqs. (\ref{tr})-(\ref{thetar}). This celestial coordinates have the same form as that for the Kerr metric, while with different $\xi$ and $\eta$. Here, we assume that the observer is located at infinity with angular coordinate $\theta_{0}$. The coordinates $\alpha$ and $\beta$ are the apparent perpendicular distances of the image as seen from the axis of symmetry and from its projection on the equatorial plane, respectively.

Supposing the observer is located in the equatorial plane of the black hole, one has $\theta_{0}=\frac{\pi}{2}$. Then the celestial coordinates $\alpha$ and $\beta$ will be of the simple form
\begin{eqnarray}
 \alpha&=&-\xi,\\
 \beta&=&\pm\sqrt{\eta}.
\end{eqnarray}
Now, the photon is assumed to be parametrized by $(\xi,\eta)$, which are conserved quantities according to the null geodesics. It is clear that this new expression of $\alpha$ and $\beta$ is independent of $\phi$, which is because that the EMDA black hole has the axisymmetry. Letting the parameters
$(\xi,\eta)$ take all possible values, then the capture region in the parameter space $(\alpha,\beta)$ will be obtained. This capture region is the shadow of the black hole, which in fact is the region in the parameter space $(\alpha,\beta)$ not illuminated by the photon sources located at infinity and distributed uniformly in all directions.

\begin{figure*}
\subfigure[]{\label{AB1a}
\includegraphics[width=6.5cm]{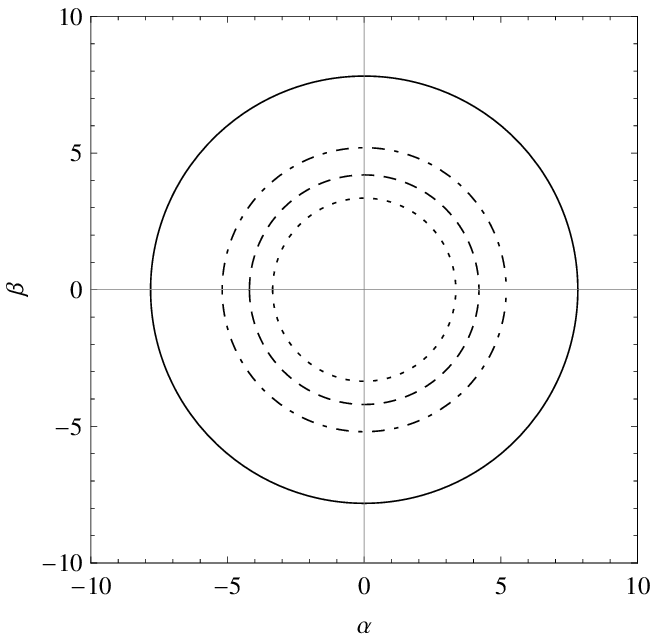}}
\subfigure[]{\label{AB1b}
\includegraphics[width=6.5cm]{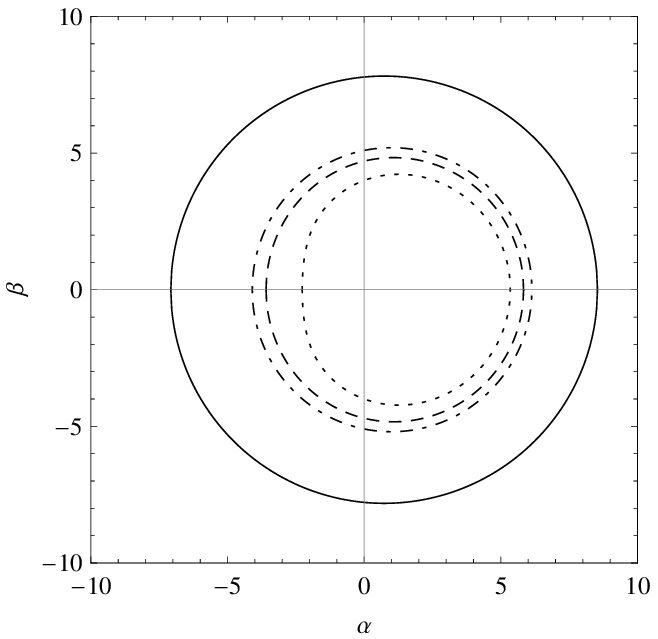}}\\
\subfigure[]{\label{AB1c}
\includegraphics[width=6.5cm]{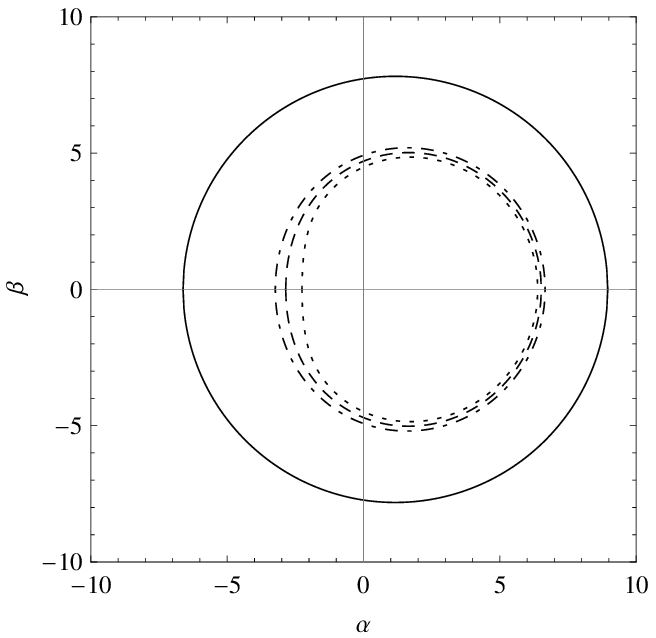}}
\subfigure[]{\label{AB1d}
\includegraphics[width=6.5cm]{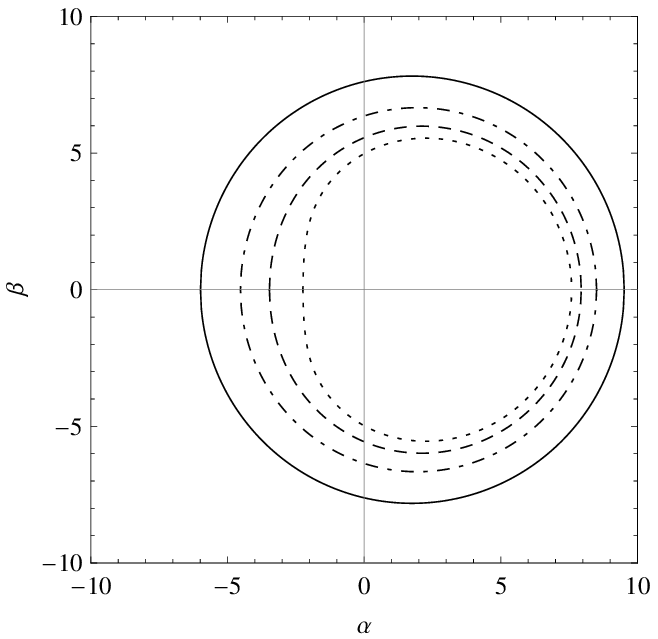}}
\caption{Shadows of EMDA black hole situated at the origin of coordinates with inclination angle $\theta_{0}=\pi/2$. (a) $a=0$, $b=-2$ (full line), 0 (dashed-dotted line), 0.5 (dashed line), and
0.8 (dotted line). (b) $a=0.5$, $b=-2$ (full line), 0 (dashed-dotted line), 0.2 (dashed line), and 0.49 (dotted line). (c) $a=0.8$, $b=-2$ (full line), 0 (dashed-dotted line), 0.1 (dashed line), and 0.19 (dotted line). (d) $a=1.2$, $b=-2$ (full line), -1 (dashed-dotted line), -0.5 (dashed line), and -0.21 (dotted line).} \label{AB}
\end{figure*}

For the nonrotating case $a=0$, the shadow is determined by the radius of the photon sphere. With the help of Eq. (\ref{nonxieta}), we can easily get
\begin{eqnarray}
 \alpha^{2}+\beta^{2}=\frac{r_{0}^{3}+4br_{0}^{2}+4b^{2}r_{0}}{r_{0}+2b-2}.
\end{eqnarray}
This implies that the shadow in the parameter $(\alpha,\beta)$ space is a circle with radius $\sqrt{\frac{r_{0}^{3}+4br_{0}^{2}+4b^{2}r_{0}}{r_{0}+2b-2}}$. This case is plotted in Fig. \ref{AB1a}. From it, we find, for different values of $b$, the radius of the circle is different. When $b=0$, the radius is $3\sqrt{3}$. It is also clear that, the radius decreases with $b$.

For the rotating case $a\neq 0$, the shadow in the parameter $(\alpha,\beta)$ space will not be a standard circle but a deformed one, which can be found in Figs. \ref{AB1b}-\ref{AB1d}. These figures share a property that, the size of the shadow reduces with the parameter $b$ for a fixed spin $a$. Moreover, when the black hole approaches the extremal case, the shadow will be distorted more away from a circle. With the spin $a$ increases, the shadow will shift to the right. However, this can not be observed from astronomical observation.

In order to extract the information of an astronomical object from this shadow, one must construct the observables, which could be directly observed from the astronomical observation. Here, we adopt the two observables defined in Ref. \cite{Maeda}. The first one is the radius $R_{s}$, which is defined as the radius of a reference circle passing by three points: the top position $(\alpha_{t}, \beta_{t})$ of the shadow, the bottom position $(\alpha_{b}, \beta_{b})$ of the shadow, and the point $(\alpha_{r}, 0)$ corresponding to the unstable retrograde circular orbit seen from an observer located in the equatorial plane. Another observable is the distortion parameter $\delta_{s}$ defined by $D/R_{s}$. Here, $D$ is the difference between the endpoints of the circle and of the shadow, both of them correspond to the prograde circular orbit and locate at the opposite side of the point $(\alpha_{r}, 0)$. Among these two observalbes, $R_{s}$ measures the approximate size of the shadow, and $\delta_{s}$ gives its deformation with respect to the reference circle. With the know of the inclination angle $\theta_{0}$, and precise enough measurements of $R_{s}$ and $\delta_{s}$, one can get the values of the spin parameter $a$ and dilaton parameter $b$ adimensionalized with the ADM mass $M_{ADM}$ of the black hole.

\begin{figure*}
\subfigure[]{\label{Rs0}
\includegraphics[width=6.5cm]{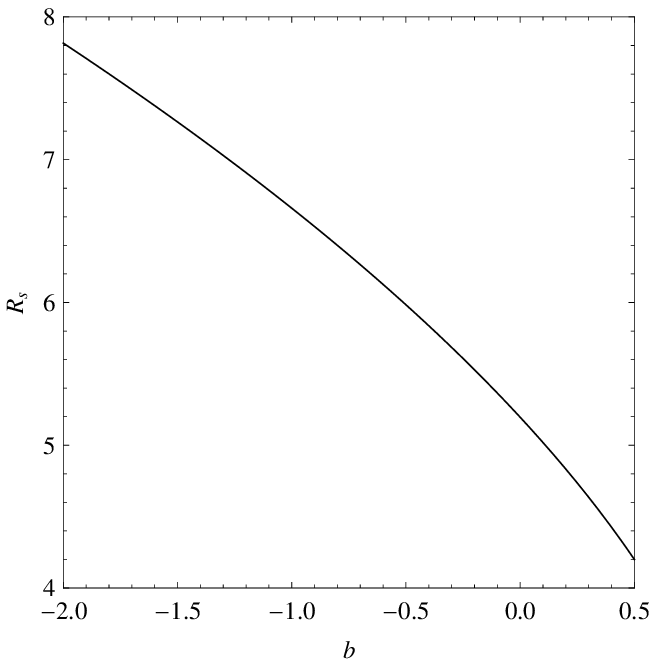}}
\subfigure[]{\label{Deltas}
\includegraphics[width=6.64cm]{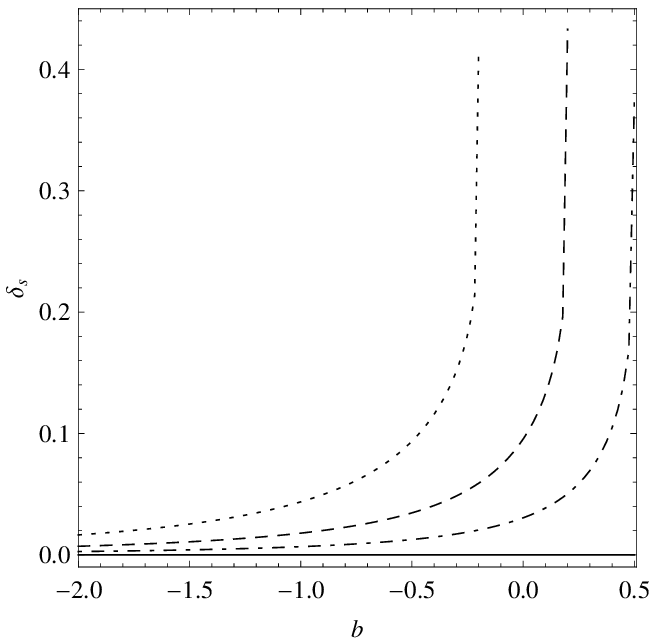}}
\caption{Behaviors of observables $R_{s}$ and $\delta_{s}$ as function of the parameter $b$. (a) a=0. (b) a=0 (full line), 0.5 (dashed-dotted line), 0.8 (dashed line), and 1.2 (dotted line).} \label{RsDeltas}
\end{figure*}

\begin{figure*}
\centerline{\includegraphics[width=6.5cm]{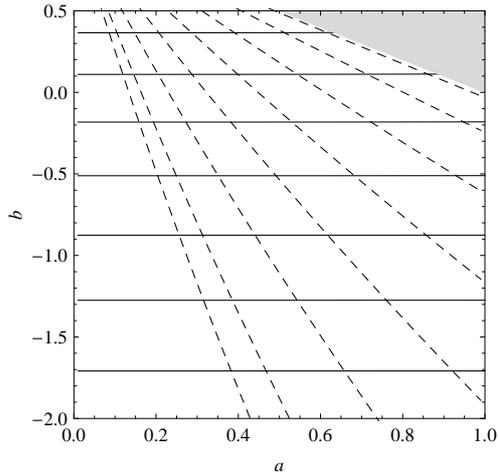}}
\caption{Contour plots of the observables $R_{s}$ and $\delta_{s}$ in the plane $(a, b)$. $R_{s}$ is described by the full lines with values 4.5, 5, 5.5, 6, 6.5, 7, and 7.5 from top to bottom. $\delta_{s}$ is described by the dashed lines with values 0.002, 0.003, 0.006, 0.012, 0.024, 0.048, 0.096, and 0.192 from left to right. The light gray zone represents naked singularities.} \label{Rs}
\end{figure*}

From the geometry of the shadow, the observable $R_{s}$ can be expressed in the form
\begin{eqnarray}
 R_{s}=\frac{(\alpha_{t}-\alpha_{r})^{2}+\beta_{t}^{2}}{2(\alpha_{r}-\alpha_{t})},
 \label{observable1}
\end{eqnarray}
where the relations $\alpha_{b}=\alpha_{t}$ and $\beta_{b}=-\beta_{t}$ are used. And the observable $\delta_{s}$ is
\begin{eqnarray}
 \delta_{s}=\frac{(\tilde{\alpha}_{p}-\alpha_{p})}{R_{s}}.
 \label{observable2}
\end{eqnarray}
$(\alpha_{p}, 0)$ and $(\tilde{\alpha}_{p}, 0)$ are the points where the contour of the shadow and the reference circle cut the horizontal axis at the opposite side of $(\alpha_{r}, 0)$, respectively. Consider the relation $\tilde{\alpha}_{p}=\alpha_{r}-2R_{s}$, the parameter $\delta_{s}$ can be further expressed as
\begin{eqnarray}
 \delta_{s}=2-\frac{D_{s}}{R_{s}},\label{delt}
\end{eqnarray}
where $D_{s}=\alpha_{r}-\alpha_{p}$ is just the diameter of the shadow along $\beta=0$. Form Eq. (\ref{delt}), it is easy to find that smaller $D_{s}$ and larger $R_{s}$ will produce large $\delta_{s}$. If $D_{s}=2R_{s}$, which corresponds to the nonrotating case, we get $\delta_{s}=0$, which implies that the shadow of the nonrotating case has no deformation as we expected. In Fig. \ref{Rs0}, the observable $R_{s}$ is shown as a function of the dilaton parameter $b$ for the nonrotating case. From it, we find that $R_{s}$ decreases with the parameter $b$. Since $R_{s}$ measures the approximate size of the shadow, positive values of $b$ generate a decrease in the size of the shadow, while negative values generate an enlargement one. After some calculations, one could obtain the results that the behavior of $R_{s}$ is almost the same for different values of spin $a$. The detailed behavior is listed in Table \ref{tab1}. It is easy to find that, for a fixed value of $b$, the difference of $R_{s}$ between $a=0$ and $a=1.2$ is of order $10^{-4}\sim 10^{-3}$, which causes a small variation in the size of the shadow. This result is similar to the Kerr black hole case and the braneworld black hole case \cite{Maeda,AmarillaEiroa,Amarilla13}. In Fig. \ref{Deltas}, the observable $\delta_{s}$ is plotted as a function of the dilaton parameter $b$ for different values of the spin $a$ of the black hole, i.e., a=0 (full line), 0.5 (dashed-dotted line), 0.8 (dashed line), and 1.2 (dotted line). We find that for different parameter $b$, $\delta_{s}$ always keeps null for $a=0$ (nonrotating case), which means the contour of the shadow is a perfect circle with no deformation. For other values of the spin $a$, it is a monotonically increasing curve and the distortion $\delta_{s}$ takes its maximal when the black hole approaches to the extremal case. The shadow of the black hole will approach to its reference circle when $b$ gets more negative. For a fixed dilaton parameter $b$, the distortion of the shadow increases with the spin $a$. The contour curves of constant $R_{s}$ and $\delta_{s}$ in the plane $(a, b)$ are plotted in Fig. \ref{Rs}. From it, we see that the contour curves of constant $R_{s}$ are almost horizontal lines. And each point is characterized by four values, i.e, $a$, $b$, $R_{s}$, and $\delta_{s}$. If fixing $R_{s}$ and $\delta_{s}$ from observations, then we are allowed to directly read out the spin $a$ and dilaton parameter $b$ of the black hole through the point in the Fig. \ref{Rs}, where the contour curves of constant $R_{s}$ and $\delta_{s}$ intersect.

\begin{table}[h]
\begin{center}
\begin{tabular}{|cc|c|c|c|c|c|c|}
  \hline
  $a$ && $b$=-2  & $b$=-1.5 & $b$=-1.0 & $b$=-0.5 & $b$=0   & $b$=0.5 \\\hline\hline
  0   && 7.81634 & 7.26452  & 6.66038  & 5.98380  & 5.19615 & 4.19960 \\\hline
  0.5 && 7.81635 & 7.26453  & 6.66041  & 5.98388  & 5.19648 & 4.20245 \\\hline
  0.8 && 7.81637 & 7.26459  & 6.66055  & 5.98426  & 5.19792 & $\cdot\cdot\cdot$ \\\hline
  1.2 && 7.81650 & 7.26485  & 6.66112  & 5.98577  & $\cdot\cdot\cdot$     & $\cdot\cdot\cdot$ \\
  \hline
\end{tabular}
\caption{Observable $R_{s}$ for different values of the dilaton paramere $b$ and spin $a$ of the shadow for the EMDA  black hole situated at the origin of coordinates with the inclination angle $\theta_{0}=\frac{\pi}{2}$.}\label{tab1}
\end{center}
\end{table}


Here, we would like to give a brief summary for the black hole shadow.
For a fixed value of spin $a$, the presence of a positive (negative) dilaton parameter leads to a larger (smaller) shadow and a less (more) distorted shadow than that of the Kerr black hole. For a fixed value of the dilaton parameter $b$, the shadow is almost the same. And there is a more distorted shadow when the black hole approaches an extremal one.

\begin{figure*}
\subfigure[]{
\includegraphics[width=6.5cm]{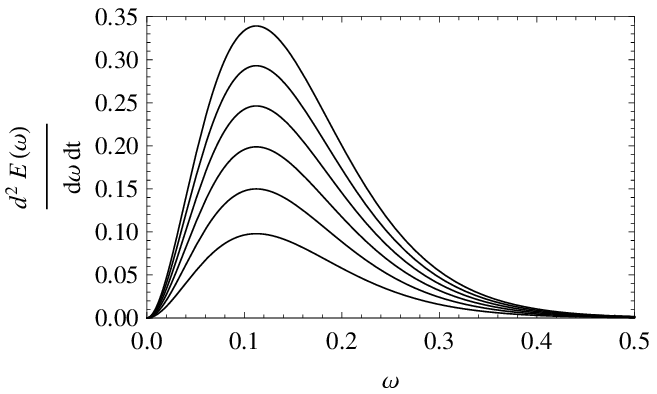}}
\subfigure[]{
\includegraphics[width=6.64cm]{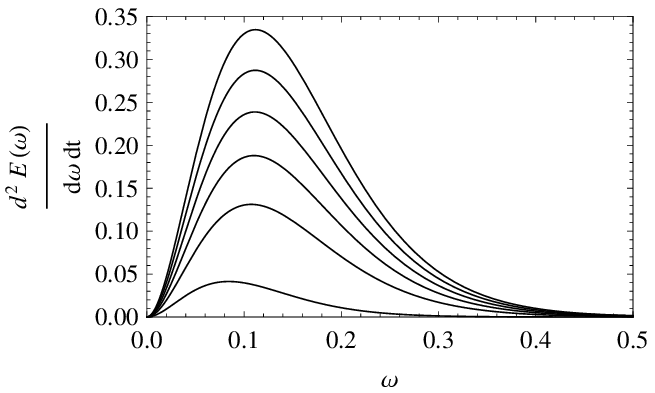}}
\caption{Behaviors of the energy emission rate $\frac{d^{2}E(\omega)}{d\omega dt}$ for $b=$-2, -1.5, -1, -0.5, 0 , and 0.5 from top to bottom. (a): a=0. (b): a=0.4.} \label{ENR}
\end{figure*}

On the other hand, we conjecture that the black hole shadow corresponds to its high-energy absorption cross section for the observer located at infinity. In general, the absorption cross section oscillates around a limiting constant value $\sigma_{\text{lim}}$ for a spherically symmetric black hole. $\sigma_{\text{lim}}$ was found to be equal to the geometrical cross section of its photon sphere \cite{Mashhoon,Misner}, while the fluctuations around the limiting value were also studied in \cite{Decanini}. Since the shadow measures the optical appearance of a black hole, it in fact is equal to the limiting constant value of the high-energy absorption cross section. It is also natural to extended this result to these rotating black holes. From Fig. \ref{Deltas}, we see that the shadow is almost a circle even for a near extremal black hole. Therefore, the limiting constant value $\sigma_{\text{lim}}$ can be approximately expressed as
\begin{eqnarray}
 \sigma_{\text{lim}}\approx \pi R_{s}^{2}.
\end{eqnarray}
With the help of $\sigma_{\text{lim}}$, the energy emission rate of the black hole in the high energy case reads
\begin{eqnarray}
 \frac{d^{2}E(\omega)}{d\omega dt}=\frac{2\pi^{3}R_{s}^{2}}{e^{\omega/T}-1}\omega^{3}
\end{eqnarray}
with the Hawking temperature $T=\frac{r_{+}^{2}-a^{2}}{4\pi r_{+}(r_{+}^{2}+2 b r_{+}+a^{2})}$. In Fig. \ref{ENR}, we show the energy emission rate against the frequency $\omega$ for $a=0$ and $a=0.4$, respectively. From it, one can see that there exists a peak of the energy emission rate for the black hole. When the dilaton parameter $b$ increases, the peak decreases and shifts to the low frequency.

\section{Naked singularity shadow}
\label{NSS}

The presence of naked singularities will lead to the violation of the cosmic censorship hypothesis \cite{Penrose}. And near a singularity, physical laws or even GR will break down. Thus, naked singularities are full of paradoxes. However, in such case, one could think that the central singularity is replaced by some high curvature region due to some quantum gravity effects. On the other hand, this hypothesis has so far not yet been proven, so naked singularities may exist in nature. Therefore, it is worth to study the naked singularity shadow and we will study it briefly in this section.

From Eq. (\ref{horizon}), it is easy to get that, when $1-a<b<1+a$ (range of light grey in Fig. \ref{BHR}), the horizon will fade out, and a naked singularity appears at the origin. Thus, the shadow of a naked singularity will be very different from that of a black hole one. As pointed out in \cite{Maeda}, for a nonrotating naked singularity, the shadow contains two parts, a dark circumference and a dark point at the center, which correspond to the photon sphere and the singularity. And other zone inside the dark circumference is not dark, which is because that the photons near both sides of the circumference will come back to the observer due to the nonexistence of the horizon. And, for a rotating naked singularity, instead of a dark circumference, there is an open arc caused by the unstable spherical photon orbits with a positive radius. The finishing points of the arc $(\alpha_{l},\pm\beta_{l})$ can be obtained form Eqs. (\ref{alpha}) and (\ref{beta}). If the observer is above the equatorial plane, he or she will see a dark spot due to the unstable spherical photon orbits with a negative radius. And if the observer is on the equatorial plane, the dark spot will shrink to a dark point with a straight line connecting it and the dark arc. In realistic observation, the neighborhood of the open-arc should be darkened. Thus, the shadow may appear as a dark ``lunate''. In the following, we will study the shadow of an EMDA naked singularity. With constructing new observables, we will discuss how to determine the spin parameter $a$ and dilaton parameter $b$ for a naked singularity when the observer is located in the equatorial plane.

\begin{figure*}
\subfigure[]{\label{NAB4a}
\includegraphics[width=6.5cm]{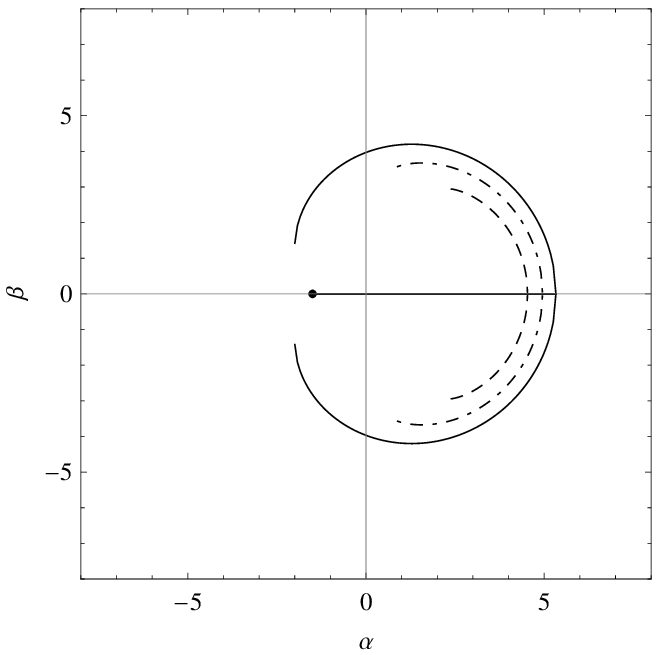}}
\subfigure[]{\label{NAB4d}
\includegraphics[width=6.5cm]{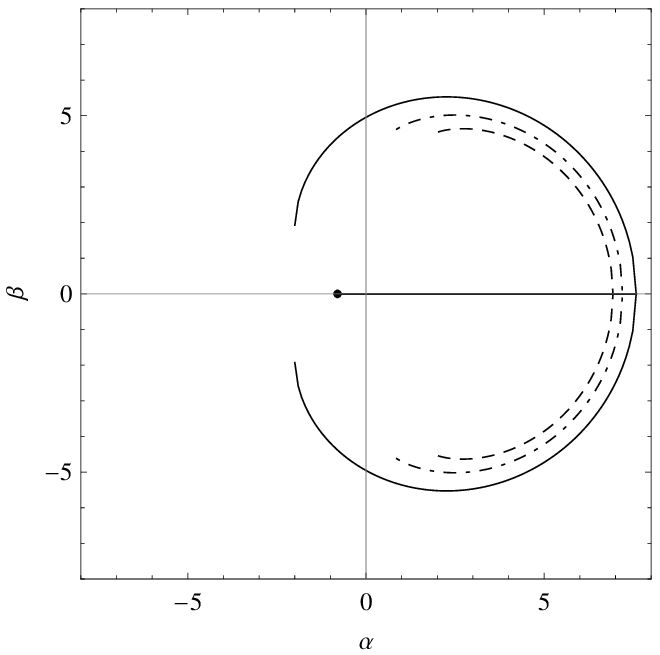}}
\caption{Shadows of naked singularity with inclination angle $\theta_{0}=\pi/2$. (a) $a=0.5$, $b=0.50001$ (full line), 0.7 (dashed-dotted line), and 0.9 (dashed line). (b) $a=1.2$, $b=-0.19999$ (full line), 0.1 (dashed-dotted line), and 0.3 (dashed line).} \label{NAB}
\end{figure*}

In Fig. \ref{NAB}, we show the shadows of the rotating EMDA naked singularity with inclination angle $\theta_{0}=\pi/2$ for  $a=0.5$ and 1.2, respectively. Form these figures, we clear see that the shadow is a dark arc plus a dark line which connects the arc and the dark point. And it is obvious that the arc of the shadow tends to closed for small dilaton parameter $b$ and tends to open for large value of $b$. The radius of the shadow is also found to increase with the spin $a$.

\begin{figure*}
\subfigure[]{\label{NRs}
\includegraphics[width=6.5cm]{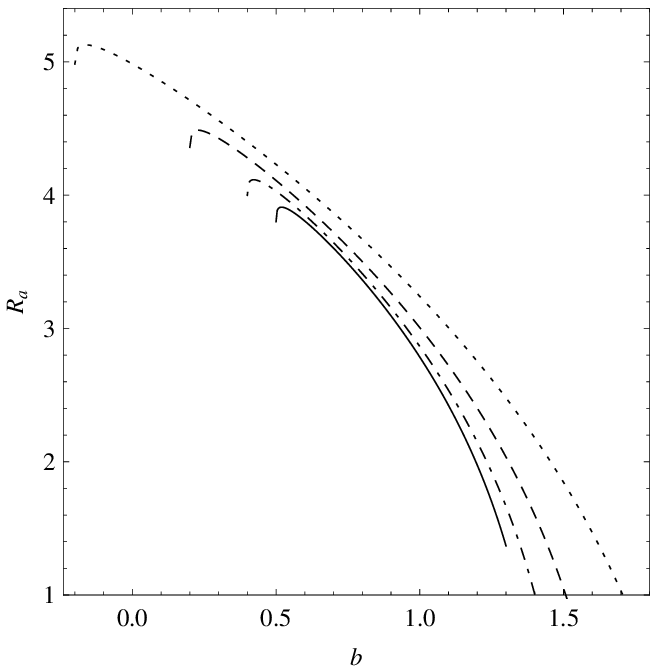}}
\subfigure[]{\label{Nphia}
\includegraphics[width=6.64cm]{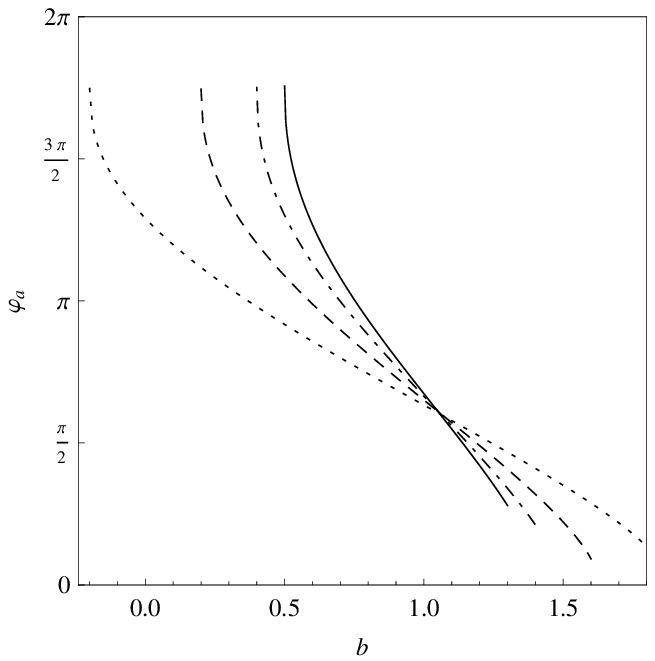}}
\caption{Behaviors of observables $R_{a}$ and $\varphi_{a}$ as function of the parameter $b$. The spin is set to $a$=0.5 (full line), 0.6 (dashed-dotted line), 0.8 (dashed line), and 1.2 (dotted line).} \label{NRsNphia}
\end{figure*}

As mentioned above, the shadow of a rotating naked singularity, which consists of a dark arc and a dark spot or a dark straight line, is very different from a rotating black hole one. So, the two observables defined in Eqs. (\ref{observable1}) and (\ref{observable2}) are no longer valid for this case. However, another two new observables \cite{Maeda} can be adopted to describe the shadow of a naked singularity. The first one is the radius $R_{a}$ defined as the radius of the circumference passing by the middle point $(\alpha_{r},0)$ and the two points $(\alpha_{l},\pm\beta_{l})$, where the arc stops:
\begin{eqnarray}
 R_{a}=\frac{(\alpha_{l}-\alpha_{r})^{2}+\beta_{l}^{2}}{2(\alpha_{r}-\alpha_{l})},
\end{eqnarray}
and the other one is $\varphi_{a}$, defined by the angle subtended by the arc, seen from the center of the circumference used to define $R_{a}$,
\begin{eqnarray}
 \varphi_{a}=2\arctan\bigg(\frac{\beta_{l}}{\alpha_{l}+R_{a}-\alpha_{r}}\bigg).
\end{eqnarray}
The behavior of the observable $R_{a}$ is shown in Fig. \ref{NRs} as a function of the dilaton parameter $b$ for several values of the spin $a$ of the naked singularity: a=0.5 (full line), 0.6 (dashed-dotted line), 0.8 (dashed line), and 1.2 (dotted line). For different vales of the spin $a$, the behavior is similar. As the dilaton parameter $b$ increases, $R_{a}$ has a sudden increase to its maximum, and then monotonically decreases with $b$. As pointed out in Ref. \cite{AmarillaEiroa}, the initial growth is related to the particular shape of the circumference of reference. And the decrease ends when the arc is generated. For a fixed value of the dilaton parameter $b$, the radius $R_{a}$ increases with the spin $a$. The angle $\varphi_{a}$ subtended by the end points of the shadow, measured from the center of the circumference of reference, is presented in Fig. \ref{Nphia} as a function of $b$ for a=0.5, 0.6, 0.8, and 1.2, respectively. These curves are monotonically decreases with $b$, and all values of $\varphi_{a}$ are limited in the range $(0,2\pi)$. We can also find that all these curves have an approximate intersection at $b=1.05$. For fixed $b\leq 1.05$, the angle decreases with the spin $a$, while for $b\geq 1.05$, it has an opposite behavior.

\section{Discussion}
\label{Summary}

In this paper, we study the shadow of the rotating EMDA black hole. With the help of the null geodesic, the celestial coordinates $\alpha$ and $\beta$ are obtained with the suppositions that the black hole is situated at the origin of coordinate and the observer is located in the equatorial plane ($\theta_{0}=\pi/2$). Then the visualization of black hole shadow is plotted in this celestial coordinates. Through constructing two observables $R_{s}$ and $\delta_{s}$, that one describes the approximate size of the black hole shadow and another measures its deformation, the shadow is examined in detail for the different values of the dilaton parameter $b$ and spin $a$. The results show that, for a fixed value of the spin $a$, the size of the shadow described by $R_{s}$ decreases with the dilaton parameter $b$. It is also found to be larger than that of Kerr case for the negative value of $b$ and smaller for positive value. The observable $\delta_{s}$ monotonically increases with the dilaton parameter $b$ and takes its maximal when the black hole approaches to the extremal case. For a fixed value of $b$, the larger value of spin $a$ is, the more the shadow distorts. The standard contour curves of constant $R_{s}$ and $\delta_{s}$ in the plane $(a, b)$ are shown in Fig. \ref{Rs}. With the observations of $R_{s}$ and $\delta_{s}$, we are allowed to read out the spin $a$ and dilaton parameter $b$ of the black hole from this figure. Based on the assumption that the area of the black hole shadow equals to the high-energy absorption cross section, we express the energy emission rate with $R_{s}$. From the behavior of energy emission rate, we find that when the dilaton parameter $b$ increases, the peak decreases and shifts to the low frequency.

We also study the shadow of the EMDA naked singularities. The results are similar to the Kerr naked singularities \cite{Maeda}, where the shadow has two parts, the dark arc and the dark spot or the straight line. Since the arc is one-dimensional, it is hard to observe with astronomical instruments. However, the neighborhood of the arc, in a realistic scenario, will also be darkened. Then, instead of a dark arc, there may be a dark ``lunate'' shadow, and it is more easily observed for the large value of $\varphi_{a}$. Therefore, we still have opportunities to observe the shadow of the naked singularities in spite of the difficulty from astronomical observations in the future. The observables $R_{a}$ and $\varphi_{a}$ proposed in \cite{Maeda} are calculated for the shadow of the naked singularities.

\begin{table}[h]
\begin{tabular}{|c||c|c|c||c|c|c||c|c|c|}
  \hline
  &\multicolumn{2}{c}{$a=0$}&&\multicolumn{2}{c}{$a=0.8$}&&\multicolumn{2}{c}{$a=1.2$}&\\\hline
     & $b$=-2 & $b$=-0.5 & $b$=0 & $b$=-2 & $b$=-0.5 & $b$=0 & $b$=-2 & $b$=-1 & $b$=-0.5\\
\hline\hline
  $R_{s}$($\mu$as)& 7.816 & 5.984 & 5.196 & 7.816 & 5.984 & 5.198 & 7.817 & 6.661 & 5.986 \\\hline
  $\delta_{s}$($\%$)  & 0 & 0 & 0 & 0.7 & 3.5 & 9.5 & 1.6 & 4.4 & 9.4 \\\hline
  $\theta_{s}$($\mu$as)& 39.972 & 30.600 & 26.573 & 39.972 & 30.603 & 26.582 & 39.973 & 34.064 & 30.611 \\
  \hline
\end{tabular}
\caption{The observables $R_{s}$, $\delta_{s}$, and the angular radius $\theta_{s}$ for the supermassive black hole Sgr A$^{*}$ at the center of the Milky Way.}\label{tab2}
\end{table}

For a approximatively estimation, the angular radius of the shadow can be calculated by using the observable $R_{s}$ as $\theta_{s}=R_{s}M/D_{O}$ with $D_{O}$ the distance between the observer and the black hole. For an arbitrary black hole of mass $M$ and distance $D_{O}$ from the observer, the angular radius can be expressed as $\theta_{s}=9.87098\times 10^{-6}R_{s}(M/M_{\odot})(1\text{kpc}/D_{O})$$\mu$as \cite{AmarillaEiroa} with $M_{\odot}$ the mass of the sun. For an example, we consider the supermassive black hole Sgr A$^{*}$ located at the Galactic center. Its mass is estimated to be $M=4.3\times 10^{6}M_{\odot}$. Setting the observer on the earth lying in the equatorial plane of the black hole, then the distance $D_{O}=8.3$ kpc \cite{Guillessen}. Thus, the observables $R_{s}$, $\delta_{s}$, and the angular radius $\theta_{s}$ are obtained and shown in Table \ref{tab2}. According to the result, the resolution of 1 $\mu$as will be enough to extract the information of the dilaton parameter $b$ from further observations, while for the spin $a$, the resolution of less than 1 $\mu$as is needed. These are out of the capacity of the current astronomical observations but can be likely to be observed with the Event Horizon Telescope at wavelengths around 1mm based on VLBI, and with the space-based radio telescopes RadioAstron \cite{w0,w1}.

\section*{Acknowledgement}
This work was supported by the National Natural Science Foundation of China (Grant No. 11205074, Grant No. 11075065, and Grant No. 11375075), the Huo Ying-Dong Education Foundation of the Chinese Ministry of Education (Grant No. 121106), and the Fundamental Research Funds for the Central Universities (Grant No. lzujbky-2013-21 and No. lzujbky-2013-18).


\begin{thebibliography}{99}


\bibitem{Eisenhauer}
  F. Eisenhauer \emph{et al}.,
{\em SINFONI in the Galactic Center: young stars and IR flares in the central light month},
      \emph{Astrophys. J.} \textbf{628}, 246 (2005),
              [arXiv:astro-ph/0502129].

\bibitem{Tanaka}
  Y. Tanaka, K. Nandra, A. C. Fabian, H. Inoue, C. Otani, T. Dotani, K. Hayashida, K. Iwasawa, T. Kii, H. Kunieda, F. Makino and M. Matsuoka,
   {\em Gravitationally redshifted emission implying an accretion disk and massive black hole in the active galaxy MCG-6-30-15},
    \emph{Nature (London)} \textbf{375}, 659 (1995);
  C. S. Reynolds and M. A. Nowak,
    {\em Fluorescent iron lines as a probe of astrophysical black hole systems},
    \emph{Phys. Rep.} \textbf{377}, 389 (2003), [arXiv:astro-ph/0212065];
  J. E. McClintock, R. Narayan, and J. F. Steiner,
    {\em Black Hole Spin via Continuum Fitting and the Role of Spin in Powering Transient Jets}, [arXiv:1303.1583[astro-ph.HE]];
  J. E. McClintock, R. Narayan, S. W. Davis, L. Gou, A. Kulkarni, J. A. Orosz, R. F. Penna, R. A. Remillard, and J. F. Steiner,
    {\em Measuring the Spins of Accreting Black Holes},
      \emph{Class. Quant. Grav.} \emph{28}, 114009 (2011), [arXiv:1101.0811[astro-ph.HE]].



\bibitem{Darwin}
  C. Darwin,
   {\em The Gravity Field of a Particle},
  \emph{Proc. R. Soc.} \textbf{A 249}, 180 (1959).

\bibitem{BozzaCapozziello}
  V. Bozza, S. Capozziello, G. Iovane, and G. Scarpetta,
    {\em Strong field limit of black hole gravitational lensing},
    \emph{Gen. Relativ. Gravit.} \textbf{33}, 1535 (2001), [arXiv:gr-qc/0102068];
  V. Bozza,
   {\em Gravitational lensing in the strong field limit},
    \emph{Phys. Rev.} \textbf{D 66}, 103001 (2002), [arXiv:gr-qc/0208075];
  V. Bozza,
    {\em Quasi-Equatorial Gravitational Lensing by Spinning Black Holes in the Strong Field Limit},
    \emph{Phys. Rev.} \textbf{D 67}, 103006 (2003), [arXiv:gr-qc/0210109];
  V. Bozza, F. DeLuca, G. Scarpetta and M. Sereno,
    {\em Analytic Kerr black hole lensing for equatorial observers in the strong deflection limit},
    \emph{Phys. Rev.} \textbf{D 72}, 083003 (2005), [arXiv:gr-qc/0507137];
  V. Bozza, F. DeLuca, and G. Scarpetta,
   {\em Kerr black hole lensing for generic observers in the strong deflection limit},
    \emph{Phys. Rev.} \textbf{D 74}, 063001 (2006), [arXiv:gr-qc/0604093].


\bibitem{Eiroa}
  E. F. Eiroa, G. E. Romero, and D. F. Torres,
   {\em Reissner-Nordstrom black hole lensing},
    \emph{Phys. Rev.} \textbf{D 66}, 024010 (2002), [arXiv:gr-qc/0203049];
  E. F. Eiroa,
   {\em Braneworld black hole gravitational lens: Strong field limit analysis},
    \emph{Phys. Rev.} \textbf{D 71}, 083010 (2005), [arXiv:gr-qc/0410128].

\bibitem{Virbhadra}
  K. S. Virbhadra and G. F. R. Ellis,
    {\em Schwarzschild black hole lensing},
     \emph{Phys. Rev.} \textbf{D 62}, 084003 (2000), [arXiv:astro-ph/9904193];
  K. S. Virbhadra and G. F. R. Ellis,
    {\em Gravitational Lensing By Naked Singularities},
     \emph{Phys. Rev.} \textbf{D 65}, 103004 (2002);
  K. S. Virbhadra and C. R. Keeton,
   {\em Time delay and magnification centroid due to gravitational lensing by black holes and naked singularities }
     \emph{Phys. Rev.} \textbf{D 77}, 124014 (2008), [arXiv:0710.2333 [gr-qc]];
  K. S. Virbhadra,
   {\em Relativistic images of Schwarzschild black hole lensing},
     \emph{Phys. Rev.} \textbf{D 79}, 083004 (2009), [arXiv:0810.2109 [gr-qc]].

\bibitem{Chen}
  S. Chen and J. Jing,
   {\em Strong field gravitational lensing in the deformed Horava-Lifshitz black hole},
   \emph{Phys. Rev.} \textbf{D 80}, 024036 (2009), [arXiv:0905.2055[gr-qc]];
 C. Ding, C. Chen, S. Kang, and J. Jing,
{\em Strong field gravitational lensing in the noncommutative black-hole spacetime},
  \emph{Phys. Rev.} \textbf{D 83} (2011) 084005,
       [arXiv:1012.1670[gr-qc]].

\bibitem{SarkarBhadra}
    K. Sarkar and A. Bhadra,
{\em Strong field gravitational lensing in scalar tensor theories},
       \emph{Class. Quant. Grav.} \textbf{23} (2006) 6101,
              [arXiv:gr-qc/0602087].

\bibitem{Whisker}
  R. Whisker,
{\em Strong gravitational lensing by braneworld black holes},
          \emph{Phys. Rev.} \textbf{D 71} (2005) 064004,
       [arXiv:astro-ph/0411786].

\bibitem{Bhadra}
  A. Bhadra,
{\em Gravitational lensing by a charged black hole of string theory},
          \emph{Phys. Rev.} \textbf{D 67} (2003) 103009,
       [arXiv:gr-qc/0306016].

\bibitem{Vazquez}
  S. E. Vazquez and E. P. Esteban,
{\em Strong field gravitational lensing by a Kerr black hole},
          \emph{Nuovo Cim.} \textbf{B 119} (2004) 489,
       [arXiv:gr-qc/0308023].

\bibitem{Nandi}
  K. K. Nandi, Y. Z. Zhang and A. V. Zakharov,
{\em Gravitational lensing by wormholes},
          \emph{Phys. Rev.} \textbf{D 74} (2006) 024020,
       [arXiv:gr-qc/0602062].

\bibitem{Gyulchev75}
  G. N. Gyulchev and S. S. Yazadjiev,
{\em Kerr-Sen dilaton-axion black hole lensing in the strong deflection limit},
           \emph{Phys. Rev.} \textbf{D 75} (2007) 023006,
       [arXiv: gr-qc/0611110];
  G. N. Gyulchev and S. S. Yazadjiev,
{\em Gravitational Lensing by Rotating Naked Singularities},
           \emph{Phys. Rev.} \textbf{D 78}, (2008) 083004,
       [arXiv:0806.3289[gr-qc]].

\bibitem{wei}
  S.-W. Wei and Y. X. Liu,
    {\em Equatorial and quasi-equatorial gravitational lensing by Kerr black hole pierced by a cosmic string},
   \emph{Phys. Rev.} \textbf{D 85}, 064044 (2012), [arXiv:1107.3023 [hep-th]].

\bibitem{Bardeen}
  J. Bardeen, {\em Black Holes}, $\acute{E}$cole d'$\acute{e}$t$\acute{e}$ de Physique Th$\acute{e}$orique, Les Houches, 1972, edited by C. De Witt and
   B. S. De Witt, (Gordon and Breach Science Publishers, New York, 1973).

\bibitem{Chandrasekhar}
  S. Chandrasekhar,
  {\em The Mathematical Theory of Black Holes}
    (Oxford University Press, New York, 1992).

\bibitem{Falcke}
  H. Falcke, F. Melia, and E. Agol,
   {\em Viewing the Shadow of the Black Hole at the Galactic Center},
    \emph{Astrophys. J.} \textbf{528}, L13 (2000), [ 	arXiv:astro-ph/9912263].

\bibitem{Vries}
 A. de Vries,
  {\em Deep Westerbork 1.4 GHz Imaging of the Bootes Field},
  \emph{Class. Quant. Grav.} \textbf{17}, 123 (2000).


\bibitem{Takahashi}
  R. Takahashi,
   {\em Shapes and Positions of Black Hole Shadows in Accretion Disks and Spin Parameters of Black Holes},
    \emph{Astrophys. J.} \textbf{611}, 996 (2004), [arXiv:astro-ph/0405099].

\bibitem{Hioki}
  K. Hioki and U. Miyamoto,
   {\em Hidden symmetries, null geodesics, and photon capture in the Sen black hole},
    \emph{Phys. Rev.} \textbf{D 78}, 044007 (2008), [arXiv:0805.3146 [gr-qc]].

\bibitem{Bambi}
  C. Bambi and K. Freese,
   {\em Apparent shape of super-spinning black holes},
     \emph{Phys. Rev.} \textbf{D 79}, 043002 (2009), [arXiv:0812.1328 [astro-ph]];
   C. Bambi and N. Yoshida, 
   {\em Shape and position of the shadow in the $\delta$=2 Tomimatsu-Sato space-time},
      \emph{Class. Quant. Grav.} \textbf{27}, 205006 (2010), [arXiv:1004.3149[gr-qc]].

\bibitem{Kraniotis}
  G. V. Kraniotis,
   {\em Precise analytic treatment of Kerr and Kerr-(anti) de Sitter black holes as gravitational lenses},
     \emph{Class. Quant. Grav.} \textbf{28}, 085021 (2011), [arXiv:1009.5189 [gr-qc]].


\bibitem{Zakharov}
   A. F. Zakharov, A. A. Nucita, F. DePaolis, and G. Ingrosso,
    {\em Direct Measurements of Black Hole Charge with Future Astrometrical Missions},
          \emph{New Astron. Rev.} \textbf{10}, 479 (2005), [arXiv:astro-ph/0505286].

\bibitem{Bozzagrg}
   V. Bozza,
    {\em Gravitational Lensing by Black Holes},
    \emph{Gen. Rel. Grav.} \textbf{42}, 2269 (2010), [arXiv:0911.2187 [gr-qc]].

\bibitem{Schee}
   J. Schee and Z. Stuchlik,
    {\em Optical phenomena in brany Kerr spacetimes},
      \emph{Int. J. Mod. Phys.} \textbf{D 18}, 983 (2009), [arXiv:0810.4445 [astro-ph]].

\bibitem{Maeda}
   K. Hioki and K. I. Maeda,
    {\em Measurement of the Kerr Spin Parameter by Observation of a Compact Object's Shadow},
    \emph{Phys. Rev.} \textbf{D 80}, 024042 (2009), [arXiv:0904.3575 [astro-ph.HE]].

\bibitem{Amarilla}
    L. Amarilla, E. F. Eiroa, and G. Giribet,
     {\em Null geodesics and shadow of a rotating black hole in extended Chern-Simons modified gravity},
      \emph{Phys. Rev.} \textbf{D 81}, 124045 (2010), [arXiv:1005.0607 [gr-qc]].

\bibitem{AmarillaEiroa}
    L. Amarilla and E. F. Eiroa,
     {\em Shadow of a rotating braneworld black hole},
  \emph{Phys. Rev.} \textbf{D 85}, 064019 (2012), [arXiv:1112.6349 [gr-qc]].


\bibitem{YumotoNitta}
    A. Yumoto, D. Nitta, T. Chiba, and N. Sugiyama,
     {\em Shadows of Multi-Black Holes: Analytic Exploration},
      \emph{Phys. Rev.} \textbf{D 86}, 103001 (2012), [arXiv:1208.0635 [gr-qc]].

\bibitem{Amarilla13}
    L. Amarilla and E. F. Eiroa,
     {\em Shadow of a Kaluza-Klein rotating dilaton black hole},
       \emph{Phys. Rev.} \textbf{D 87}, 044057 (2013), [arXiv:1301.0532 [gr-qc]].

\bibitem{Nedkova}
    P. G. Nedkova, V. Tinchev, and S. S. Yazadjiev,
     {\em The Shadow of a Rotating Traversable Wormhole},
        [arXiv:1307.7647[gr-qc]];
    V. K. Tinchev and S. S. Yazadjiev,
      {\em Possible imprints of cosmic strings in the shadows of galactic black holes},
        [arXiv:1311.1353[gr-qc]];
    Z. Li and C. Bambi,
      {\em Measuring the Kerr spin parameter of regular black holes from their shadow}, [arXiv:1309.1606[gr-qc]].


\bibitem{Garcia}
    A. Garcia, D. Galtsov, and O. Kechkin,
     {\em Class of Stationary Axisymmetric Solutions of the Einstein-Maxwell-Dilaton-Axion Field Equations},
        \emph{Phys. Rev. Lett.} \textbf{74}, 1276 (1995).

\bibitem{w0}
 http://www.eventhorizontelescope.org

\bibitem{w1}
    http://www.asc.rssi.ru/radioastron.

\bibitem{Mashhoon}
 B. Mashhoon,
  {\em Scattering of Electromagnetic Radiation from a Black Hole},
   \emph{Phys. Rev.} \textbf{D 7}, 2807 (1973).

\bibitem{Misner}
 C.W. Misner, K. S. Thorne, and J. A. Wheeler, \emph{Gravitation}
(Freeman, San Francisco, 1973).

\bibitem{Decanini}
    Y. Decanini, G. Esposito-Farese, and A. Folacci,
     {\em Universality of high-energy absorption cross sections for black holes},
       \emph{Phys. Rev.} \textbf{D 83}, 044032 (2011), [arXiv:1101.0781 [gr-qc]].


\bibitem{Penrose}
  R. Penrose,
   {\em GRAVITATIONAL COLLAPSE: THE ROLE OF GENERAL RELATIVITY},
    \emph{Riv. Nuovo Cim.} \textbf{1}, 252 (1969).

\bibitem{Guillessen}
    S. Gillessen, F. Eisenhauer, S. Trippe, T. Alexander, R. Genzel, F. Martins, and T. Ott,
     {\em Monitoring stellar orbits around the Massive Black Hole in the Galactic Center},
       \emph{Astrophys. J.} \textbf{692}, 1075 (2009), [arXiv:0810.4674 [astro-ph]].

\end{thebibliography}
\end{document}